\documentclass[aps,
  pra,
  twocolumn,  superscriptaddress,
  nofootinbib]{revtex4-2}
\usepackage{mathrsfs}

\usepackage{hyperref}
\usepackage{graphicx,amsmath,amsfonts}% Include figure files
\usepackage{amssymb}

\newcommand{\vers}[1]{\hat{\mathbf{#1}}}

\usepackage{xcolor}
\usepackage{amsfonts}
\usepackage{amsmath}
\usepackage{amssymb}
\usepackage{mathtools}

\renewcommand{\vec}[1]{\mathbf{#1}}

\newcommand{\partm}{\mu}
\newcommand{\partn}{\nu}
\newcommand{\dV}{\, \mathrm{d}\Omega}

\newcommand{\sca}{\mathrm{s}}
\newcommand{\abs}{\mathrm{a}}
\newcommand{\ext}{\mathrm{e}}
\newcommand{\inc}{\mathrm{i}}

\newcommand{\mat}{\mathrm{m}}
\newcommand{\ed}{\mathrm{ed}}
\newcommand{\param}{\boldsymbol{\vartheta}}

\newcommand{\ep}{\mathrm{ep}}
\newcommand{\mutua}{M}

\newcommand{\parity}{\mathscr{P}}
\newcommand{\trev}{\mathscr{T}}
\newcommand{\PT}{\parity\trev}
\newcommand{\dyad}[1]{\overset{\text{\scalebox{.7}{$\boldsymbol{\leftrightarrow}$}}}{\mathbf{#1}} }
\newcommand{\mean}[1]{\langle #1 \rangle}

\newcommand{\volume}{V_a}

\renewcommand{\Re}{\operatorname{Re}}
\renewcommand{\Im}{\operatorname{Im}}

\begin{document}
%\linenumbers % activate line numbering
\author{Emanuele Corsaro}
\affiliation{ Department of Electrical Engineering and Information Technology, Universit\`{a} degli Studi di Napoli Federico II, via Claudio 21, Napoli, 80125, Italy}
\author{Filippo Capolino}
\email[]{f.capolino@uci.edu}
\affiliation{Department of Electrical Engineering and Computer Science, University of California, Irvine, CA 92697 USA}
\author{Carlo Forestiere}
\email[]{carlo.forestiere@unina.it}
\affiliation{ Department of Electrical Engineering and Information Technology, Universit\`{a} degli Studi di Napoli Federico II, via Claudio 21, Napoli, 80125, Italy}
\title{Exceptional Points in the Scattering Resonances of a Sphere Dimer}

%\title{Exceptional Points of Degeneracy \\ in the Electromagnetic Scattering from a Dimer of Spheres}

% ===== Abstract (replace your current abstract) =====
\begin{abstract}
We investigate exceptional points of degeneracy (EPDs) in electromagnetic scattering of a sphere dimer from the electroquasistatic limit to the fully retarded regime. In the quasistatic limit, we prove that $\parity\trev$-symmetric configurations, realized by spheres with complex-conjugate susceptibilities, host EPDs.  Beyond this limit, retardation breaks $\mathscr{PT}$-symmetry; nevertheless, by jointly tuning the material dispersion of the two spheres, we derive analytic synthesis conditions for  realizing  EPDs at \textit{real frequencies}. Near an EPD, we show that single-parameter perturbations yield the characteristic square-root splitting of the eigenfrequencies, and we quantify its impact on scattering, extinction, and absorption, clarifying sensing implications.
\end{abstract}

\maketitle

%%%%
%\input{SEC/Introduction}
\section{Introduction}

\textit{Exceptional points of degeneracy} (EPDs) are points in the parameter space of a non-Hermitian operator, at which two or more eigenvalues and their corresponding eigenvectors simultaneously coalesce \cite{kato1966perturbation, heiss_exceptional_2004, heiss_physics_2012}. At such points, the operator becomes defective (non-diagonalizable), and the associated eigenbasis loses completeness. { Introduced by Kato} in Ref.~\cite{kato1966perturbation} in the context of perturbation theory, EPDs
have gained considerable attention in optics and photonics during the last decade \cite{feng_non-hermitian_2017, othman_exceptional_2017, el-ganainy_non-hermitian_2018, ozdemir_paritytime_2019, miri_exceptional_2019}. A salient feature of systems exhibiting EPDs is the extreme sensitivity of both eigenvalues and eigenvectors to small perturbations in system parameters. This inherent enhanced sensitivity provides a powerful mechanism for detecting very small variations in the system's parameters via the induced perturbations in the spectral properties of the system \cite{wiersig_enhancing_2014,hodaei_enhanced_2017,nikzamir_highly_2022, wiersig_review_2020}. 

{
Several studies \cite{el-ganainy_non-hermitian_2018, ozdemir_paritytime_2019, miri_exceptional_2019} have employed \textit{coupled-mode theory} (CMT) \cite{haus_coupled-mode_1991} to investigate EPDs in parity-time ($\PT$) symmetric systems. While CMT is attractive because it is conceptually simple, offers clear physical intuition, and requires minimal computational effort, it does not precisely capture the underlying physics. In particular, coupled electromagnetic resonators are intrinsically non-$\PT$ symmetric due to retardation in the radiative coupling, which leads to a complex and \textit{frequency-dependent} coupling that cannot be easily included in conventional CMT. Moreover, a recent study \cite{dmitriev_retardation-induced_2023}, employing the \textit{coupled-dipole approximation} (CDA) \cite{purcell_scattering_1973, draine_discrete-dipole_1988, draine_discrete-dipole_1994, steshenko_single_2009,meier_enhanced_1983,maier_optical_2003,forestiere_role_2009, kelly_optical_2003, yang_discrete_1995}, has shown that these retardation effects can themselves induce EPDs even in systems with no gain and loss.}

 In this work, we investigate EPDs in electromagnetic scattering by a dimer of homogeneous, linear, time-dispersive spheres with arbitrary material dispersion. By using a volume-integral formulation of Maxwell's equations, we recast the problem as an equivalent two-port circuit under the same assumptions as the coupled-dipole approximation \cite{yang_discrete_1995,kelly_optical_2003,steshenko_single_2009}. Relative to the standard CDA, our formulation provides two main advantages: it disentangles the roles of excitation, geometry, material properties, radiation losses and scattering reactive fields, and it naturally leads to a state-space formulation. These features are essential for simplifying the analytical derivation and physical interpretation of EPD conditions. We prove that, in the electroquasistatic regime, a $\parity\trev$-symmetric sphere dimer with complex-conjugate susceptibilities supports EPDs.  Beyond this limit, retardation breaks $\parity\trev$-symmetry; nevertheless, real-frequency EPDs can still emerge when the dispersions of the two materials are jointly tuned, and we derive general conditions for their occurrence.  These predictions are validated against rigorous multiparticle Mie theory calculations \cite{xu_electromagnetic_1995}. Finally, we quantify the square-root spectral splitting to single-parameter perturbations and its impact on scattering, extinction, and absorption, highlighting the potential of this platform for highly sensitive sensing applications.

%%%%
%\input{SEC/Model}
\section{The Model}
\label{sec:Model}

\begin{figure*}
     \includegraphics[width=0.9\linewidth]{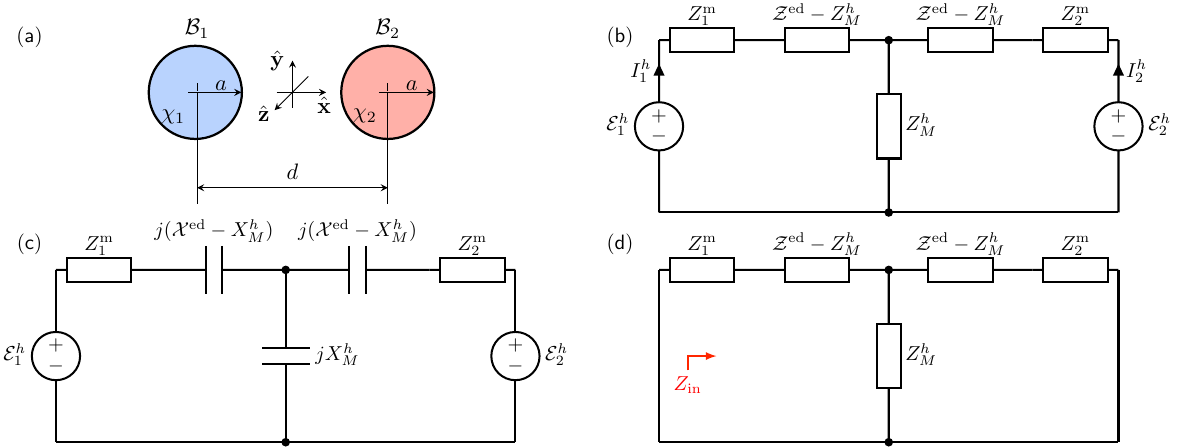}
    \caption{(a) Dimer composed of two spheres of radius $a$ and inter-center distance $d$. (b) Two-port equivalent circuit of two interacting particles in the coupled dipole approximation and in the full-wave regime. (c) Two-port equivalent circuit of two interacting particles in the coupled dipole approximation and in the static limit. (d) Configuration for characterizing the resonant mode of the dimer (i.e., $Z_{\mathrm{in}}=0$). In panels (b),(c),(d) $h\in\{x,y,z\}$ denotes the polarization of the dipolar mode.}
    \label{fig:dimer}
    \label{fig:circuit_2port}
    \label{fig:circuit_2port_qs} 
\end{figure*}
Consider a dimer composed of two spherical particles, labeled by $\partm = 1,2$, each made of a linear, homogeneous, isotropic, and time-dispersive material, characterized by an electric susceptibility $\chi_{\partm}$, embedded in vacuum. Each sphere has radius $a$ and is centered at $\vec{r}_\partm$, with the inter-center distance denoted by $d$. A time-harmonic dependence $e^{j\omega t}$ is assumed. Let $\mathcal{B}_{\partm}$ be the bounded three-dimensional domain occupied by the $\partm$-th particle, and $\partial \mathcal{B}_\partm$ its boundary. {Without loss of generality, we assume that the dimer is aligned along the $x$-axis [see Fig.~\ref{fig:dimer}(a)].}

A standard way to formulate the full-wave electromagnetic scattering problem, following the volume-integral-equation framework (see, e.g., \cite{hanson_operator_2002,mishchenko_electromagnetic_2014}), is to introduce as unknown the polarization current density \(\mathbf{J}\) in the region \(\mathcal{B}=\mathcal{B}_1\cup\mathcal{B}_2\) induced by the incident field \(\mathbf{E}^{\inc}\), which satisfies the homogeneous Helmholtz equation in vacuum.
The volume-integral formulation enables a disentanglement of material response, radiative and reactive scattered fields, and geometry \cite{forestiere_volume_2018}, crucial for deriving EPD synthesis conditions.

In the following, we adopt the notation described in Appendix \ref{sec:notations}, where for any vector field $\vec{F} = \vec{F}(\vec{r})$ we denote by  $\vec{F}_\mu$ its restriction on $\mathcal{B}_\mu$. In the region $\mathcal{B}_\partm$, the field ${\bf J}_{\partm}$ is related to the total electric field  ${\mathbf{E}}_{\partm}$, which is the sum of the scattered field ${\mathbf{E}}_{\partm}^{\sca}$ and the incident field ${\mathbf{E}}_{\partm}^{\inc}$, i.e. $\vec{E}_{\partm} = \vec{E}_{\partm}^{\inc} + \vec{E}_{\partm}^{\sca}$, by the constitutive relation
\begin{equation}
a \, Z_{\partm}^{\mat} \, \vec{J}_{\partm} \left( {\bf r} \right) = {\bf E}_{\partm}^{\sca}  \left( {\bf r} \right)+{\bf E}_{\partm}^{\inc}  \left( {\bf r} \right) \qquad  \mathbf{r} \in \mathcal{B}_\partm,
 \label{eq:constrel}
\end{equation}
where  $Z_{\partm}^{\mat}$
%$ = Z_{\partm}^{\mat} \left( \omega \right)$
is the \textit{material impedance} of the $\mu$-th particle \cite{forestiere_first-principles_2024, corsaro_mie_2026}, defined as
 \begin{equation}
    Z_{\partm}^{\mat} :=   \frac{\zeta_0}{jk_0 a}\frac{1}{\chi_\partm}.
    \label{eq:Zmat}
\end{equation}
Here, $k_0 = \omega \sqrt{\varepsilon_0 \mu_0}$,  and $\zeta_0 = \sqrt{\mu_0/\varepsilon_0}$ is the vacuum characteristic impedance, with $\varepsilon_0$ and $\mu_0$ denoting the vacuum permittivity and permeability, respectively.  The scattered field in $\mathcal{B}_\partm$ is expressed as 
\begin{equation}
    \vec{E}_{\partm}^{\sca}(\vec{r}) =  \sum_{\partn=1}^{2} \vec{E}_{\partm,\partn}^{\sca}(\vec{r}), 
    \label{eq:Esca_s}
\end{equation}
where $\vec{E}^{\sca}_{\partm,\partn}$ denotes the contribution to the scattered field on $\mathcal{B}_\partm$ arising from the current density $\vec{J}_\partn$ on the $\partn$-th particle alone. The scattered field $\vec{E}^{\sca}_{\partm,\partn}$  is related to its source $\vec{J}_\partn$ through the volume integral operator $\mathscr{L}_\partm$ as 
\begin{equation}
     \vec{E}_{\partm,\partn}^{\sca} = - a \mathscr{L}_\partm\{\vec{J}_{\partn}\} \quad \vec{}
 \label{eq:Esca_opL}
\end{equation}
where \(\mathscr{L}_\partm\) denotes the restriction of the integral operator \(\mathscr{L}\) to the volume \(\mathcal{B}_\partm\), applied to the field \(\mathbf{J}_\partn\). The operator \(\mathscr{L}\) is defined as in Eq. \eqref{eq:opLdef}.

We use the set of polarization current density modes $\{ \mathbf{j}_{\partm}^k\}_{k=1}^{\infty}$ to represent the solution of the scattering problem \cite{forestiere_first-principles_2024}:
\begin{equation}
\mathbf{J}_{\partm} = \frac{1}{a^2}  \sum_{k=1}^{\infty} I_{\partm}^k\, \vec{j}_{\partm}^k,
\label{eq:MIMexpansion}
\end{equation}
where $I_{\partm}^{k} = I_{\partm}^{k} (\omega)$ are the weights. The  polarization current density modes $\mathbf{j}^{k}_{\partm}$ are the solutions of the following eigenvalue problem in  each individual bounded domain $\mathcal{B}_\partm$:
\begin{equation}
    \mathscr{L}_\partm \left\{\mathbf{j}^{k}_{\partm}\right\} = \mathcal{Z}^{k} \, \mathbf{j}^{k}_{\partm}, \qquad  \vec{r} \in \mathcal{B}_\partm
    \label{eq:EigProb}
\end{equation}
where $\mathcal{Z}^k = \mathcal{R}^{k} + j \mathcal{X}^{k}$ is the \textit{mode impedance} of the current eigenmodes $\vec{j}_\partm^k$,  $\mathcal{R}^{k}$ is the \textit{radiation resistance}, and  $\mathcal{X}^{k}$ is the \textit{modal reactance} \cite{forestiere_first-principles_2024, corsaro_mie_2026}.  The current density modes have support within the particle volume and are, in general, not orthogonal with respect to the standard inner product on $\mathcal{B}$, defined in Eq. \eqref{eq:InnerDef}. However, they satisfy the bi-orthogonality relation \cite{bergman_theory_1980,forestiere_material-independent_2016}:
\begin{equation}
    \langle \vec{j}^{h*}_{\partm }|\vec{j}^{k}_{\partm}\rangle = N^{h}_\partm\delta_{hk},
    \label{eq:bi-orth}
\end{equation}
 where $\delta_{hk}$ denotes the Kronecker delta and $N^h_\partm = \langle \vec{j}^{h*}_{\partm }|\vec{j}^{h}_{\partm}\rangle$. Without loss of generality, we impose the normalization condition $\|\vec{j}^k_\partm\|^2 := {\langle \vec{j}^k_\partm | \vec{j}^k_\partm \rangle} = 1$ on the current density modes.
 
Combining Eqs. \eqref{eq:Esca_s}, \eqref{eq:Esca_opL}, \eqref{eq:MIMexpansion}, and \eqref{eq:EigProb}, and using the linearity of the operator $\mathscr{L}$,  Eq. \eqref{eq:constrel} is rewritten as
\begin{equation}
%\begin{cases}
%\sum_{h}Z_{11}I_{h}^{(1)}\vec{j}_{h}^{(1)}  + I_h^{(2)}\mathscr{L}^{(1)}\{\vec{j}_h^{(2)}\} = a \vec{E}_0^{(1)}, \quad \vec{r} \in \mathcal{B}_1 \\
%        \\
%\sum_{h}Z_{22}I_{h}^{(2)}\vec{j}_{h}^{(2)}+ I_h^{(1)}\mathscr{L}^{(2)}\{\vec{j}_h^{(1)}\}  = a \vec{E}_0^{(2)}, \quad \vec{r}\in \mathcal{B}_2.
%\end{cases}
\sum_{k=1}^\infty \left[ (Z^{\mat}_{\partm} + \mathcal{Z}^{k})I^{k}_{\partm} \, \vec{j}^{k}_{\partm}  + I^k_{\partn}\mathscr{L}_{\partm}\{\vec{j}^k_{\partn}\} \right] = a \vec{E}^{\inc}_{\partm}, \quad (\partm \neq \partn).
\end{equation}
Projecting along $\mathbf{j}^{h*}_{\partm}$, using \eqref{eq:bi-orth}, we obtain
\begin{equation}
       (Z^{\mat}_{\partm} + \mathcal{Z}^{h})I_{\partm}^{h} +  \sum_{k=1}^{\infty} Z^{hk}_{\partm \partn }    I^k_{\partn} = \mathcal{E}^{h}_{\partm}
         \label{eq:ModelZI} \quad (\mu \neq \nu),
\end{equation}
where $
    \mathcal{E}^{h}_{\partm} = a \langle \vec{j}^{h*}_{\partm} |\vec{E}^{\inc}_\partm \rangle/N^h_{\partm}
$
is an equivalent voltage source, and 
\begin{equation}
    Z^{hk}_{\partm \partn} := \frac{1}{N^h_\partm}\left\langle \vec{j}^{h*}_{\partm}  | \mathscr{L}_\partm \{\vec{j}^k_\partn\} \right \rangle \quad (\mu \neq \nu)
    \label{eq:Zmutua_def}
\end{equation}
is a \textit{mutual} impedance that takes into account the coupling between the two particles.

{
We now assume that the scattering response of each sphere is dominated by the electric-dipole modes, each characterized by the impedance $ \mathcal{Z}^{h}=\mathcal{Z}^{\mathrm{ed}}$, { for $h=1,2,3$},
%$h \in \mathcal{I}=\{x,y,z\}$. 
{ where $\mathcal{Z}^{\ed} = \mathcal{R}^\ed + j\mathcal{X}^\ed$ is the electric-dipole  impedance. The electric-dipole radiation resistance ($\mathcal{R}^{\ed}$) and reactance ($\mathcal{X}^\ed$) are shown as a function of $k_0a$ in Fig. \ref{fig:Z_mutua}(a)-(b), respectively.  This impedance is always ohmic–capacitive $(\mathcal{X}^{\ed} <0)$, and it can be expanded as in Eq. \eqref{eq:Zed_approx}. 
\begin{figure}[t]
    \centering
    \includegraphics[width=1\linewidth]{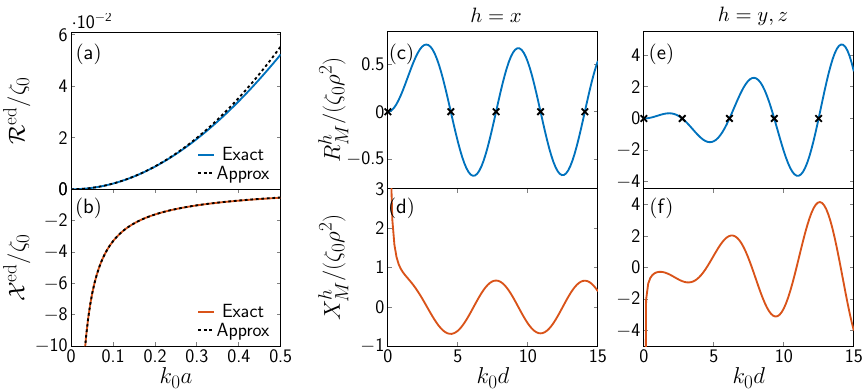}
    \caption{(a) Normalized electric-dipole resistance $\mathcal{R}^{\ed}/\zeta_0$ (blue curve) and (b) reactance $\mathcal{X}^\ed/\zeta_0$ (red curve) as a function of $k_0a$. The RLC approximations in Eq. \eqref{eq:Zed_approx} (dashed lines) are provided for reference. (c),(e) Normalized mutual resistance $R_{\mutua}^h/(\zeta_0 \rho^2)$ and (d),(f) reactance $X_{\mutua}^h/(\zeta_0 \rho^2)$ of the longitudinal (c)-(d) and transverse (e)-(f) coupling as a function of the electric distance $k_0d$. The first zeros of the mutual resistance are marked with $\times$. }
    \label{fig:Z_mutua}
\end{figure}
{
We denote by $\langle \cdot \rangle$ the integral mean over $\mathcal{B}$, as defined in Eq.~\eqref{eq:MeanDef}. Since the vectors 
$\langle \vec{j}_{\partm}^1 \rangle$, $\langle \vec{j}_{\partm}^2 \rangle$, 
and $\langle \vec{j}_{\partm}^3 \rangle$ are aligned with the Cartesian 
directions $\vers{x}$, $\vers{y}$, and $\vers{z}$, respectively, we hereafter adopt the convention $h \in \{x, y, z\}$ in place 
of $h \in \{1, 2, 3\}$.
}
}

We also assume that the electric field generated by the $\nu$-th particle is constant within $\mathcal{B}_\mu$ ($\mu \neq \nu)$. 
These approximations are valid when the particles are electrically small ($k_0a \ll 1$), {the real part of their permittivity is either negative or, if positive, not excessively large}, and when they are sufficiently well separated ($d\gg a$). Under these conditions, higher-order multipoles can be neglected, and the field generated by the $\nu$-th particle within the domain of the $\mu$-th particle can be approximated by its value at the particle center, that is, $\mathscr{L}_\partm \{\vec{j}^k_\partn\}(\vec{r}) \approx   a^{-1} V_a \zeta_0 jk_0 \dyad{G}_{\partm\partn} \cdot\mean{\vec{j}^k_\partn}$, where $\dyad{G}_{\partm\partn} = \dyad{G}(\vec{r}_\partm,\vec{r}_\partn)$ is the free-space dyadic Green function  [cf. Eq.~\eqref{eq:dyadG}] evaluated at the centers of particles $\partm$ and $\partn$}, and $V_a = \frac{4}{3}\pi a^3$. Substituting this approximation into Eq. \eqref{eq:Zmutua_def}, and using the facts that $\dyad{G}_{\partm\partn}$ is diagonal and symmetric, that $\mean{\vec{j}^h_{\partm}}\cdot\mean{\vec{j}^k_{\partn}} = \delta_{hk}$, and that $N_{\partm}^h \approx 1$ for $k_0a \ll 1$, the mutual impedances are equal to $Z^{hk}_{12} = Z^{hk}_{21} = Z_{\mutua}^h\delta_{hk}$, where
\begin{equation}
Z_{\mutua}^h = \begin{cases}
-\zeta_0 \rho^2 \, \dfrac{2}{3}\dfrac{1+jk_0d}{jk_0d}e^{-jk_0d}, &\text{if }h=x\\ \\
\zeta_0 \rho^2 \, \dfrac{1}{3}  
     \dfrac{1+jk_0d+(jk_0d)^2}{jk_0d} e^{-jk_0d} &\text{if }h=y,z
\end{cases}
\end{equation}
and $\rho = a/d$.
The expressions of the mutual resistance $R_{\mutua}^h = \Re\{Z_{\mutua}^h\}$ and reactance $X_{\mutua}^h = \Im\{Z_{\mutua}^h\}$ are given in Eq.~\eqref{eq:RXmutua} and are plotted in Fig.~\ref{fig:Z_mutua}(c)-(f) as functions of the electric interparticle distance $k_0d$, for both longitudinal  $(h=x)$ and transverse polarizations  $(h=y,z)$. The oscillations describe in-phase and counter-phase radiative coupling.

Finally, Eq. \eqref{eq:ModelZI} is rewritten  in the form of the constitutive relation of a two-port network illustrated in Fig.~\ref{fig:circuit_2port}(b) \cite{chua_linear_1987}:
\begin{equation}
    \boldsymbol{\mathcal{Z}}^h \, \vec{I}^h = \boldsymbol{\mathcal{E}}^h.
    \label{eq:E=ZI}
\end{equation}
Here, the current and excitation vectors are defined as
$
\vec{I}^h = [ I^h_{1}, I^h_{2} ]^\intercal$ and $
\boldsymbol{\mathcal{E}}^h =[ \mathcal{E}^h_{1}, \mathcal{E}^h_{2} ]^\intercal
$, and the impedance matrix \(\boldsymbol{\mathcal{Z}}^h = \boldsymbol{\mathcal{Z}}^h(\omega; \param)\) is given by
\begin{equation}
    \boldsymbol{\mathcal{Z}}^h(\omega; \param) = 
    \begin{bmatrix}
        Z_{11} & Z_{\mutua}^h \\
        Z_{\mutua}^h   & Z_{22}
    \end{bmatrix},
    \label{eq:matZ}
\end{equation}
where $\param$ denotes the parameter vector of the system, and $Z_{11} = Z^{\mat}_1 + \mathcal{Z}^{\ed}$, $Z_{22} = Z^{\mat}_2 + \mathcal{Z}^{\ed}$.

This equivalent-circuit formulation disentangles the dependencies on geometry, material, and excitation: geometry enters only through the radiation and mutual impedances;  material properties enter solely via the material impedances; and the excitation acts only through the equivalent voltage sources $\mathcal{E}^h_{\mu}$.

 As demonstrated in Appendix \ref{sec:CDA}, under these approximations, our model is equivalent to a coupled-dipole approximation,  which is widely used in the literature of scattering by small particle arrays \cite{steshenko_single_2009,meier_enhanced_1983,maier_optical_2003,forestiere_role_2009}.

Notably, as shown in Appendix \ref{sec:Asymptotic},  Eqs. \eqref{eq:Red_approx} and \eqref{eq:Rmu_approx}, the leading-order terms in the expansions of $\mathcal{R}^{\ed}$ and $R_{\mutua}^h$ are identical and approach zero in the quasistatic limit, defined by $k_0a \to 0$ and $k_0d \to 0$.  In this limit, the electric-dipole impedance and the mutual impedance reduce to their leading capacitive terms, namely $\mathcal Z^{\ed} \sim 1/(j\omega C^{\ed})$, and $Z_{\mutua}^h \sim 1/(j\omega C_{\mutua}^h)$.
Consequently, in the quasistatic limit, the circuit in Fig. \ref{fig:circuit_2port}(b) reduces to the simplified form in Fig. \ref{fig:circuit_2port_qs}(c). In this limit, the only dissipative contribution arises from the two material resistances.

In the following discussion, we restrict our attention to the longitudinal polarization of the dipolar mode (i.e., $h=x$). For clarity of notation, the superscript $h$ will be omitted hereafter.

\section{Resonances and Exceptional Points}

{
The dynamical behavior of the circuit shown in Fig.~\ref{fig:circuit_2port}(b) can be described by a linear time-invariant state-space model with state matrix $ \boldsymbol{A}$ [cf. Eq. \eqref{eq:SS2_time}]. In the complex $\omega$-plane, the eigenfrequencies are related to the eigenvalues $\{\lambda_i\}$ of $ \boldsymbol{A}$ as  $\omega_i = -j \lambda_i$. 

The input impedance is defined as the impedance observed at one port when the remaining port is terminated by a short circuit [see Fig. \ref{fig:circuit_2port}(d)]. As shown in Appendix \ref{sec:Jordan}, the zeros of the input impedance are eigenfrequencies of the system. }  The corresponding \textit{characteristic equation} is then:
{
\begin{equation}
Z_{\mathrm{in}}(\omega;\param) = \frac{\mathrm{det}(\boldsymbol{\mathcal{Z}})}{Z_{22}} = 0.
\label{eq:Zin}
\end{equation}
 For fixed $\param$, the solution of Eq. \eqref{eq:Zin} defines a discrete set of eigenfrequencies $\{\omega_{i}\}$. As $\param$ varies, the eigenfrequencies trace continuous trajectories $\omega_{i} = \omega_{i}(\param)$ in the complex plane. Our objective is to identify \textit{Exceptional Points of Degeneracy} (EPDs), namely points $\param = \param_{\ep}$ in the parameter space for which two (or more) eigenfrequencies coalesce and the corresponding eigenvectors become linearly dependent.

{
It is worth emphasizing that the mere coalescence of eigenvalues of a state matrix does not imply an exceptional point. By contrast, the presence of a double-root of $Z_{\mathrm{in}}$, implies the appearance of a nontrivial Jordan block in the resolvent and therefore the non-diagonalizability of the state matrix \cite{kailath_linear_1980} [see Appendix \ref{sec:Jordan}].
Hence, in our network formulation, requiring $\omega_{\ep}$ to be a multiple root of~\eqref{eq:Zin} provides a sufficient condition for the manifestation of an EPD.} The exceptional point $(\omega_{\ep},\param_{\ep})$ is then found as the solution of the following system:
\begin{equation}
\begin{cases}
Z_{\mathrm{in}}(\omega;\param) = 0, \\
\dfrac{\partial }{\partial \omega}Z_{\mathrm{in}}(\omega;\param) = 0 .
\label{eq:EPD_double}
\end{cases}
\end{equation}
These relations ensure that the algebraic multiplicity of $\omega_{\ep}$ is at least two. { Additionally, if the following condition is satisfied 
\begin{equation}
\det\left(\boldsymbol{\mathcal{Z}}  (\omega^*;\param)\right) = \left[ \det\left(\boldsymbol{\mathcal{Z}}(\omega;\param)\right)\right]^*,
    \label{eq:FconjSimm}
\end{equation}
then the eigenfrequencies are either real or appear in complex-conjugate pairs.
In particular, if the exceptional point originates from the coalescence of a complex-conjugate pair,
its frequency $\omega_{\ep}$ necessarily lies on the real axis.
}

As shown in Appendix \ref{sec:App_PT_symmetry}, the parity-time ($\PT$) symmetry implies the condition \eqref{eq:FconjSimm}. We recall that a system is said to be $\PT$-\textit{symmetric} if it remains invariant under the combined action of the \textit{parity} ($\parity$) and \textit{time-reversal} ($\trev$) operators \cite{bender_real_1998}, i.e.
$\PT\{\boldsymbol{\mathcal{Z}}\} = \boldsymbol{\mathcal{Z}}$. The operators $\parity$ and $\trev$ are defined in Eq. \eqref{eq:PTop_def}. In this context, the parity operator $\parity$ effectively exchanges the ports of the two-port network, while the time-reversal operator $\trev$ inverts the sign of the resistive components. 
}

%\input{SEC/PT_Symmetry}
% \subsection{Parity-Time Symmetry}
\label{sec:PT_symmetry}
{
In the quasistatic limit, the circuit in Fig. \ref{fig:circuit_2port_qs}(c) exhibits $\PT$-symmetry if the condition
\begin{equation}
    {Z}_{\Sigma}:= Z_{11} + Z_{22}^* = 0
    \label{eq:PT_condition}
\end{equation}
is satisfied.  In the quasistatic limit, condition \eqref{eq:PT_condition} is satisfied if and only if the material impedances obey $Z^{\mat}_1 =- Z^{\mat *}_{2}$, which corresponds to a dimer with complex-conjugate susceptibilities.

%We now demonstrate that, even in the absence of $\PT$-symmetry, the system can still exhibit either purely real eigenfrequencies or complex-conjugate pairs.  This reveals a more general class of behavior extending beyond the strictly $\PT$-symmetric case discussed previously. 

However, in the full-wave regime, $\PT$-symmetry of the circuit in Fig. \ref{fig:circuit_2port}(b) is generally broken because $R_M \neq 0$ almost everywhere.  Thus, in this regime, condition \eqref{eq:PT_condition} no longer guarantees the validity of Eq.~\eqref{eq:FconjSimm}. Nevertheless, it is easy to show that, if the following relation holds, 
\begin{equation}
     Z_{\Sigma}(\omega; \param) = \left(R_{\mutua} + 2j X_{\mutua} \right)\frac{R_{\mutua}}{Z_{22}},
    \label{eq:Zadd}
\end{equation}
then the characteristic equation~\eqref{eq:Zin} can be recast as
\begin{equation}
\left|Z_{22}(\omega;\param)\right|^2 - X_{\mutua}^2(\omega;\param) = 0,
\label{eq:CharEq_2}
\end{equation}
which, by construction, satisfies Eq.~\eqref{eq:FconjSimm} and therefore admits either purely real eigenfrequencies or complex-conjugate pairs.

Notably, condition~\eqref{eq:Zadd} is fulfilled when the material impedances satisfy
\begin{equation}
    Z^{\mat}_1= - Z^{\mat *}_{2} +  Z_{\Sigma} - 2 \mathcal{R}^\ed.
    \label{eq:Zmat_noPT}
\end{equation}
}
 Equation~\eqref{eq:Zmat_noPT} represents one of the central results of this work: for a given geometry and material impedance $Z^{\mat}_2$, it provides an explicit synthesis condition for $Z^{\mat}_1$ which can be directly recast in terms of the susceptibilities \(\chi_1\) and \(\chi_2\) through Eq.~\eqref{eq:Zmat}.

As a special case, at the discrete values of $k_0d$ where the mutual resistance vanishes (marked by $\times$ in Fig.~\ref{fig:Z_mutua}(c),(e)), Eq.~\eqref{eq:Zadd} gives $Z_{\Sigma}=0$. Consequently, \eqref{eq:Zmat_noPT} gives $R^{\mat}_{1} = - (R^{\mat}_{2} + 2 \mathcal{R}^{\ed})$ and $X^{\mat}_{1} = X^{\mat}_{2}$.
%%%%%%%%%%%
%\input{SEC/Results}
\section{Results}
\label{sec:results}
So far we have made no assumption about the material dispersion relation of the two spheres. In this section, we consider two spheres of radius $a$, whose susceptibility follows a Drude model, given by \cite{kreibig_optical_1995}:
\begin{equation}
\chi_\partm(\omega) = - \frac{\omega_{\mathrm{p}\partm}^2 }{\omega(\omega - j \gamma_\partm)},
\label{eq:Drude}
\end{equation}
where $\omega_{\mathrm{p} \partm} \in \mathbb{R}$ is the plasma frequency, and $\gamma_\partm \in \mathbb{R}$ is the damping (or gain) rate associated with the free electron response in the $\partm$-th particle. For the Drude susceptibility \eqref{eq:Drude}, the material impedance corresponds to a simple RL series circuit; that is, $Z^{\mat}_{\partm}(\omega) = R^{\mat}_{\partm} + j\omega L^{\mat}_{\partm}$, with $L^{\mat}_{\partm} = (\omega_{\mathrm{p}\partm}^2 \varepsilon_0 a)^{-1} $ and $R^{\mat}_{\partm} = \gamma_\partm L^{\mat}_{\partm}$. Therefore, it can resonate with the capacitive impedances of the dipole modes of the two spheres. { For the Drude susceptibility, the parameters are $\param = (\gamma_{1},\gamma_{2},k_{\mathrm{p} 1}a,k_{\mathrm{p}2}a,d/a)$ where $k_{\mathrm{p}\partm} = \omega_{\mathrm{p}\partm} \sqrt{\varepsilon_0 \mu_0}$.}

\subsection{Exceptional Points}
\begin{figure}[t]
    \centering
    \includegraphics[width=0.95\linewidth]{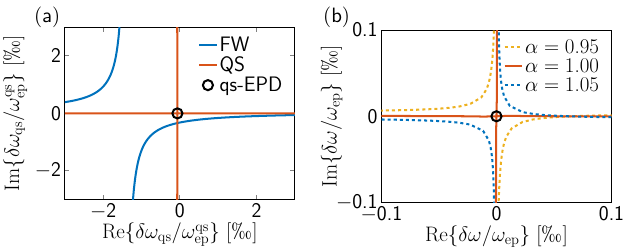}
    \caption{(a) Normalized eigenfrequency deviation 
${\delta\omega_{\mathrm{qs}}}/{\omega_{\mathrm{ep}}^{\mathrm{qs}}} 
= {(\omega_\pm - \omega_{\mathrm{ep}}^{\mathrm{qs}})}/{\omega_{\mathrm{ep}}^{\mathrm{qs}}}$ 
in the complex-frequency plane for varying $\gamma_{2} = \gamma = -\gamma_{1}$, comparing full-wave (FW, blue) and quasistatic (QS, red) models for a conjugate-symmetric ($\chi_{1} = \chi_{2}^*$) dimer of Drude spheres with $k_{\mathrm{p}1}a = k_{\mathrm{p}2}a = 0.1$ and $d = 5a$. 
(b) Normalized eigenfrequency deviations in the FW regime ${\delta\omega}/{\omega_{\mathrm{ep}}} 
= {(\omega_\pm - \omega_{\mathrm{ep}})}/{\omega_{\mathrm{ep}}}$ 
for varying $\gamma_{2}$ with $\gamma_{1} = -\gamma_{2} + \alpha\Gamma_\ep$ in a Drude sphere dimer with $k_{\mathrm{p}2}a = 0.1$, $k_{\mathrm{p}1}a = k_{\mathrm{p}1,\mathrm{ep}}a$, and $d = 5a$. Results for $\alpha \in \{0.95, 1.00, 1.05\}$ show eigenfrequency coalescence only at $\alpha = 1$. The EPD is marked with $\circ$.}
    \label{fig:ComplexFreqPlane}
    \label{fig:diffeig}
\end{figure}
As discussed in Sec.~\ref{sec:PT_symmetry}, $\PT$-symmetry can be realized in the quasistatic limit by enforcing  \eqref{eq:PT_condition}. For the Drude dispersions, this corresponds to particles with identical plasma frequencies, $\omega_{\mathrm{p}1} = \omega_{\mathrm{p}2} = \omega_\mathrm{p}$, and damping rates of equal magnitude and opposite sign, $\gamma_1 = -\gamma$ and $\gamma_2 = \gamma$, with $\gamma>0$.
In this configuration, the characteristic equation \eqref{eq:Zin} reduces to

%\begin{equation*}
%\left(\frac{\omega}{\omega_0}\right)^4 + \omega_0^2 (R^\mat)^2 C^{\ed} \left(\frac{\omega}{\omega_0}\right)^2 + 1 - \left(\frac{C^{\ed}}{C_{\mutua}}\right)^2 = 0,
%\label{eq:wstatic_circuit}
%\end{equation*}
%%
\begin{equation}
\left(\frac{\omega}{\omega_0}\right)^4 + \left[(\omega_0 R^\mat C^{\ed})^2-2 \right]\left(\frac{\omega}{\omega_0}\right)^2 + 1 - \left(\frac{C^{\ed}}{C_{\mutua}}\right)^2 = 0,
\label{eq:wstatic_circuit}
\end{equation}

where $\omega_0 = \omega_\mathrm{p} / \sqrt{3}$ is the isolated-sphere electric-dipole resonance frequency; $C^{\ed}$ and $C_{\mutua}$ are the electric-dipole and mutual capacitances, defined in Eqs. \eqref{eq:ED_lowfreq_param} and \eqref{eq:Zmu_low_freq_param}, respectively. { For the longitudinal polarization}, Eq. \eqref{eq:wstatic_circuit} is rewritten  as
\begin{equation}
\left(\frac{\omega}{\omega_0}\right)^4 
+ \left( \hat{\gamma}^2 - 2 \right)\left(\frac{\omega}{\omega_0}\right)^2 
+ 1 - 4 \rho^6 = 0,
\label{eq:wstatic}
\end{equation}
where \(\hat{\gamma} = \gamma / \omega_0\). Equation \eqref{eq:wstatic} is bi-quadratic in \((\omega/\omega_0)\), hence eigenfrequencies appear in pairs \(\pm(\omega/\omega_0)\). Since the coefficients are real, any complex roots appear as conjugate pairs. The physically relevant roots with positive real part are
\begin{equation}
\omega_{\pm}(\hat \gamma,\rho) = \omega_0 \sqrt{1 - \frac{\hat{\gamma}^2}{2} \pm \frac{1}{2} \sqrt{\hat{\gamma}^4 - 4\hat{\gamma}^2 + 16 \rho^6}},
\label{eq:w_12}
\end{equation}
{where $\sqrt{\cdot}$ is the principal square root.} Eq. \eqref{eq:w_12} reveals an exceptional point of degeneracy (EPD) in the quasistatic limit, hereafter referred to as the \textit{quasistatic Exceptional Point} (qs-EPD), obtained by nulling the inner discriminant in Eq. \eqref{eq:w_12}:
\begin{equation}
 \hat{\gamma}^4 - 4\hat{\gamma}^2 + 16 \rho^6 = 0.
    \label{eq:gamma_epd_eq}
\end{equation}
Equation~\eqref{eq:gamma_epd_eq} defines a one-dimensional manifold in the two-parameter space $(\hat{\gamma},\rho)$: only a single parameter needs to be actively tuned to reach an EPD.
Solving Eq.~\eqref{eq:gamma_epd_eq} for \(\hat{\gamma}\), we obtain
\begin{equation}
    \gamma_{\ep}^{\mathrm{qs}}(\rho) =  \, \omega_0\sqrt{2 \pm 2\sqrt{1-4 \rho^6}}.
    \label{eq:gamma_qsep}
\end{equation}
In the absence of damping/gain (\(\hat{\gamma} = 0\)), Eq.~\eqref{eq:gamma_epd_eq} is satisfied only by the trivial uncoupled limit \(\rho = 0\) consisting of widely separated particles that do not have an EPD.  Alternatively, Eq.~\eqref{eq:gamma_epd_eq} is solved for \(\rho\), providing the critical geometric condition necessary to realize an EPD for a given value of damping/gain. Notably, Eq.~\eqref{eq:gamma_epd_eq} admits two real solutions for 
$\hat{\gamma}^2$
for any physically admissible value of $\rho$. 
Indeed, the discriminant is 
$\Delta = 16(1 - 4\rho^6)$, which remains strictly positive under the 
non-overlapping condition $d > 2a$ (i.e., $\rho < 1/2$).  In what follows, we focus on the minus-sign branch of Eq.~\eqref{eq:gamma_qsep}, which corresponds to the solution with the smallest gain/loss level.
In the full-wave regime, the condition $\chi_1 = \chi_2^*$ no longer guarantees a $\PT$-symmetric configuration because of retardation and radiation losses, as discussed in Sec. \ref{sec:PT_symmetry}. In particular, if $\gamma$ is the only control parameter and all other parameters are fixed, the conjugate-symmetric choice $\chi_1=\chi_2^*$ does not produce an EPD: the eigenfrequency trajectories in the complex plane do not coalesce as $\gamma$ varies, as illustrated in Fig.~\ref{fig:ComplexFreqPlane}(a).

\begin{figure}[t]
    \centering
    \includegraphics[width=0.7\linewidth]{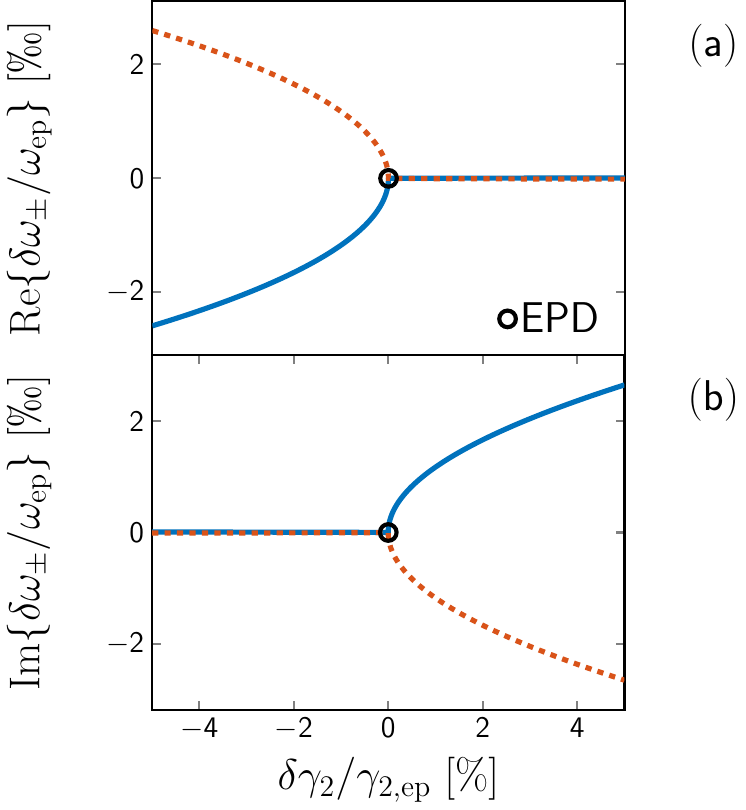}
    \caption{(a) Real and (b) imaginary parts of the normalized eigenfrequency deviations $\delta \omega_{\pm}/\omega_{\mathrm{ep}}$ of the dimer ($k_{\mathrm{p}2} a = 0.1$, $d=5a$) as a function of the normalized damping ratio detuning $\delta \gamma_2/\gamma_{2,\ep} = (\gamma_2-\gamma_{2,\ep})/\gamma_{2,\ep}$. The gain is tuned as $\gamma_{1} = -\gamma_{2} + \Gamma_\ep$. The EPD is marked with $\circ$. }
    \label{fig:eigval}
\end{figure}

{
To synthesize an exceptional point in the full-wave regime, we fix 
$k_{\mathrm{p}2} = k_{\mathrm{p}}$ and $\rho = a/d$. 
Upon imposing Eq.~\eqref{eq:Zadd}, we determine \((\omega_{\ep},\gamma_{2,\ep})\) by requiring Eq.~\eqref{eq:CharEq_2}
and its derivative with respect to \(\omega\) to vanish simultaneously. The system is solved numerically by employing a root-finding algorithm, using 
the quasistatic solution $(\omega_{\ep}^{\mathrm{qs}}, \gamma_{\ep}^{\mathrm{qs}})$ 
as the initial guess. Using Eq.~\eqref{eq:Zmat_noPT}, we then compute the 
effective resistance and inductance at the exceptional point, as $R_{1,\ep}^{\mat} = \Re\!\left\{ Z^{\mat}_1(\omega_{\ep}; 
\param_{\ep})\right\} 
$, and $L_{1,\ep}^{\mat} = {\Im\!\left\{ Z^{\mat}_1(\omega_{\ep}; 
\param_{\ep})\right\}}/{\omega_{\ep}},
$
which directly determine the Drude parameters of the gain particle as
\begin{equation}
\omega_{\mathrm{p}1,\ep} = \frac{1}{\sqrt{L_{1,\ep}^{\mat} \, \varepsilon_0 a}}, 
\qquad 
\gamma_{1,\ep} = \frac{R_{1,\ep}^{\mat}}{L_{1,\ep}^{\mat}} .
\label{eq:gain_mat}
\end{equation}
{It is important to emphasize that, unlike in the quasistatic limit, the quantity $\Gamma_\ep = \gamma_{1,\ep} + \gamma_{2,\ep}$ does not vanish}. This construction yields an exceptional point of degeneracy at $\omega = \omega_{\ep}$, as shown in Fig. \ref{fig:ComplexFreqPlane}(b). Here, the normalized eigenfrequency deviation of the dimer ($k_{\mathrm{p}2}a = 0.1$, $d=5a$)
$\delta\omega_\pm/\omega_{\ep} = (\omega_\pm - \omega_{\ep})/\omega_{\ep}$ is plotted 
in the complex-frequency plane by varying $\gamma_{2}$ together with $\gamma_1$ enforcing the constraint $\gamma_1 + \gamma_2 = \alpha \Gamma_{\ep}$. The eigenfrequencies are obtained by numerically solving the characteristic equation \eqref{eq:Zin}. The plasma frequency of the gain particle is fixed at its EPD value, i.e. $k_{\mathrm{p}1}a = k_{\mathrm{p}1,\ep}a$. Results are shown for $\alpha \in \{0.95, 1.00, 1.05\}$. 
The plot highlights that the eigenfrequencies coalesce only for $\alpha = 1$, where $\gamma_{1} = \gamma_{1,\ep}$ when 
$\gamma_{2} = \gamma_{2,\ep}$, while for $\alpha \neq 1$ the eigenfrequencies 
remain distinct and no coalescence is observed.

For the case $\alpha = 1$, Fig.~\ref{fig:eigval} shows the normalized eigenfrequency deviation $\delta \omega_\pm/\omega_{\ep}$ of the dimer as a 
function of the normalized damping ratio detuning 
$\delta \gamma_2/\gamma_{2,\ep}$, with the gain tuned according to 
$\gamma_1 + \gamma_2 = \Gamma_\ep$. The eigenfrequencies coalesce at 
$(\omega_{\ep}, \gamma_{2,\ep})$ and exhibit the characteristic square-root 
splitting away from the EPD, a hallmark of exceptional point degeneracies.

%%%%%%%%%%%%%%%%%%%%%%%%%%%%%%%%%%%%%

\subsection{Optical efficiencies in the proximity of EPDs}
\begin{figure}[t!]
    \centering
    \includegraphics[width=0.65\linewidth]{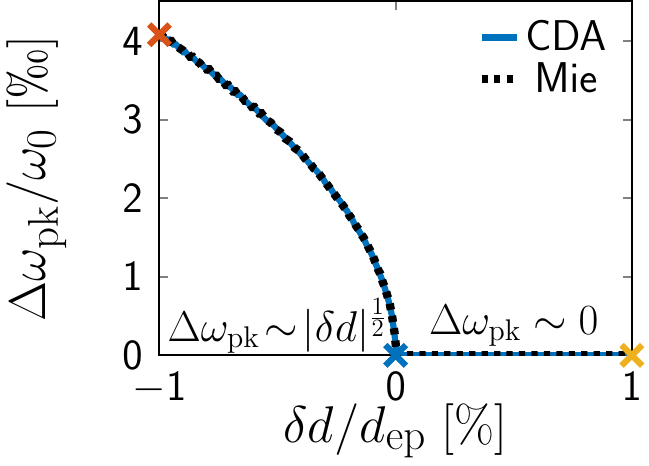}
    \caption{Peak-frequency splitting \(\Delta\omega_{\mathrm{pk}}/\omega_{0}\) versus normalized detuning \(\delta d/d_{\mathrm{ep}}\) for the geometric control parameter \(\vartheta_i=d/a\). Solid: CDA circuit model; dashed: multiparticle Mie theory. The values $d = 0.99d_{\ep}$ (red $\times$), $d = d_{\ep}$ (blue $\times$), and $d = 1.01d_{\ep}$ (yellow $\times$) are provided for reference.}
\label{fig:Sensitivity}
\end{figure}

{
To validate the model, we compare the scattering (\(Q_{\sca}\)), absorption (\(Q_{\abs}\)), and extinction (\(Q_{\ext}\)) efficiencies obtained according to our coupled-dipole circuit model against the multiparticle Mie theory \cite{xu_electromagnetic_1995}. The formulas used to compute these quantities are given in Eq. \eqref{eq:CrossSec} of Appendix \ref{sec:ScatteringAppendix}.

Specifically, we quantify the sensitivity of the scattering efficiency to single-parameter perturbations in the vicinity of the EPD. Let the dimer be tuned to $(\omega_{\mathrm{ep}},\boldsymbol{\vartheta}_{\mathrm{ep}})$ and perturb a single control parameter, holding all others at their EPD values. As shown in  Appendix~\ref{sec:Sensitivity},  the sensitivity of the system’s response, quantified here by the frequency splitting $\Delta \omega_{\mathrm{r}}:=\Re \{\omega_{+} - \omega_{-}\}$, diverges as the EPD is approached.  The frequency-splitting behavior reveals two distinct regimes, depending on the sign of the perturbation $\delta\vartheta_i$. For negative perturbations $\delta\vartheta_i < 0$, the system has two distinct resonance frequencies, and the splitting exhibits the characteristic square-root scaling $\Delta \omega_{\mathrm{r}} \sim |\delta\vartheta_i|^{1/2}$. Conversely, for $\delta\vartheta_i \geq 0$, the two resonance frequencies coalesce; in this regime, the difference manifests {predominantly} in the imaginary parts of the eigenfrequencies, associated with peak broadening, while the frequency splitting vanishes, i.e. $\Delta \omega_{\mathrm{r}}\approx 0$.

This theoretical prediction is verified in Fig.~\ref{fig:Sensitivity}. 
In this example,  we consider a plasmonic dimer composed of two Drude spheres of radius \(a\), separated by an inter-center distance \(d = 5a\). The system is illuminated by a plane wave with angular frequency $\omega$, polarized along the dimer axis and propagating along the direction perpendicular to the dimer plane. The lossy sphere is characterized by \(k_{\mathrm{p}2} a = 0.1\) and a damping rate \(\gamma_2=\gamma_{2,\ep}\), while the gain sphere is tuned according to $\gamma_1 = \gamma_{1,\ep}$ and $k_{\mathrm{p}1} = k_{\mathrm{p}1,\ep}$. Starting from the configuration $\boldsymbol{\vartheta} = \boldsymbol{\vartheta}_{\mathrm{ep}}$, we vary solely the interparticle distance (i.e. $\vartheta_i= d/a$) around the EPD value $d_{\mathrm{ep}} = 5a$, and compute the peak-frequency splitting  $\Delta \omega_\mathrm{pk} = \omega_{\mathrm{pk}+} - \omega_{\mathrm{pk}-}$, where $\omega_{\mathrm{pk}+}$ and $\omega_{\mathrm{pk}-}$ are the high- and low-frequency peaks evaluated on the efficiency spectra (e.g. Fig. \ref{fig:CrossSections}), as a function of the normalized detuning $\delta d/d_{\mathrm{ep}} = (d- d_{\ep})/d_{\ep} $. We compare the predictions of the coupled-dipole circuit model (solid line), with those obtained from the full-wave multiparticle Mie theory (dashed line), finding excellent agreement between the two approaches. Both numerical results clearly exhibit the predicted square-root behavior. This analysis confirms the characteristic square-root scaling also for the peak-frequency splitting $\Delta \omega_{\mathrm{pk}} \sim |\delta d|^{1/2}$. 

\begin{figure}[t!]
  \centering
  \includegraphics[width=0.8\linewidth]{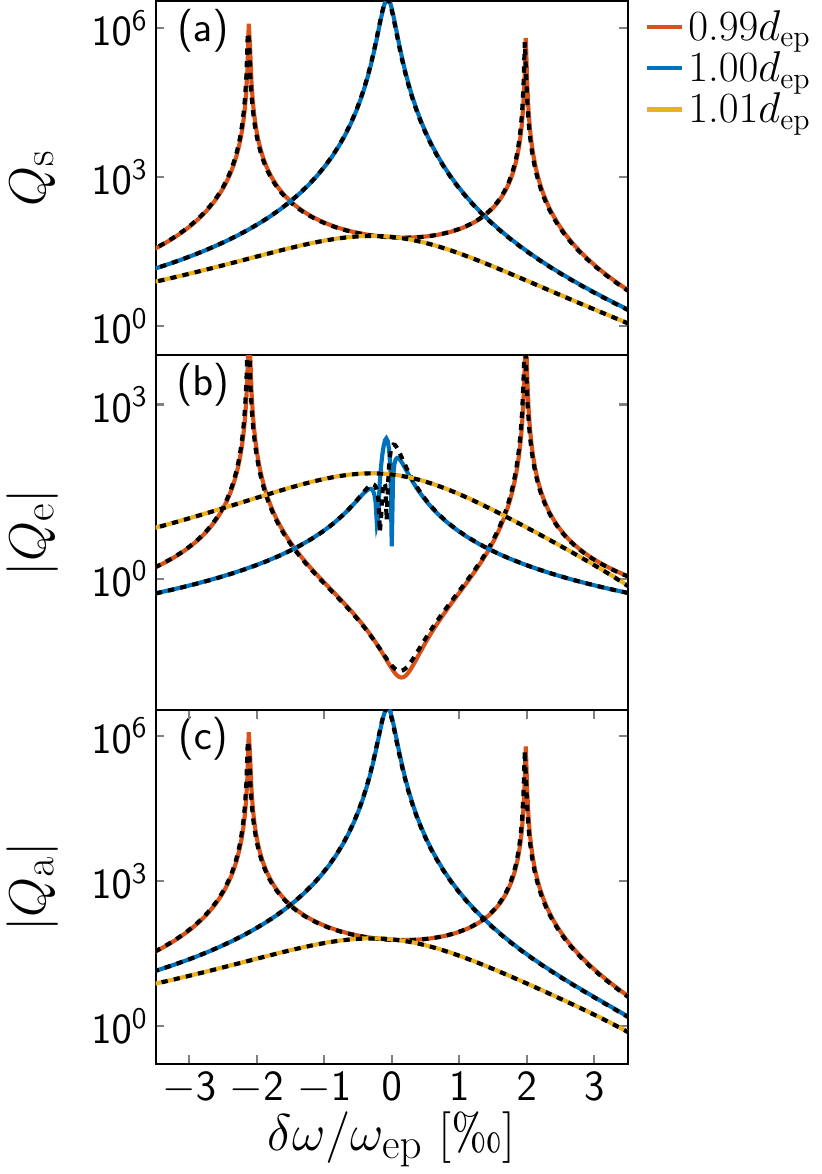}
\caption{ (a) Scattering (\(Q_{\sca}\)), (b) extinction (\(Q_{\ext}\)), and (c) absorption (\(Q_{\abs}\)) efficiencies of a Drude–sphere dimer (\(k_{\mathrm{p}2} a=0.1\), \(d=5a\)) versus normalized frequency detuning \(\delta\omega/\omega_{\mathrm{ep}}=(\omega-\omega_{\mathrm{ep}})/\omega_{\mathrm{ep}}\), for the three distances marked by $\times$ in Fig. \ref{fig:Sensitivity}, namely: \(d=0.99\, d_{\mathrm{ep}}\) (red), \(d=d_{\mathrm{ep}}\) (blue), and \(d=1.01\,d_{\mathrm{ep}}\) (yellow). Here $d_\mathrm{ep} = 5a$.  Solid curves: coupled-dipole circuit model; dashed curves: multiparticle Mie theory.}
\label{fig:CrossSections}
\end{figure}

Additionally, for three representative values of $d/d_{\ep} \in \{0.99,1.00,1.01\}$ indicated by $\times$ in Fig. \ref{fig:Sensitivity},  Figure \ref{fig:CrossSections} shows the three efficiencies as functions of the normalized frequency detuning
$\delta\omega/\omega_{\mathrm{ep}}
      = {(\omega-\omega_{\mathrm{ep}})}/{\omega_{\mathrm{ep}}}
$. The EPD transition is clear in Fig.~\ref{fig:CrossSections}. For \(d = 0.99\,d_{\mathrm{ep}}\) (red curve), two well-separated resonances appear, associated with eigenfrequencies with distinct real parts. At the critical value \(d = d_{\mathrm{ep}}\) (blue curve), the two resonances coalesce into a single peak  with a strongly enhanced amplitude, signaling coalescence of the eigenfrequencies. For \(d = 1.01\,d_{\mathrm{ep}}\) (yellow curve), the splitting is in the imaginary parts of the eigenfrequencies, and the response reduces to a single broadened resonance with reduced amplitude due to finite imaginary parts. In all three scenarios considered, $|Q_{\ext}|$ is several orders of magnitude smaller than $Q_{\sca}$ and $|Q_{\abs}|$, due to the interplay between gain and loss. The results are in good agreement with the multiparticle Mie theory (dashed line), across all cases, confirming the accuracy of the proposed coupled-dipole circuit model.}

\section{Conclusion}
We have developed a first-principles framework to predict and interpret exceptional points of degeneracy (EPDs) in the scattering resonances of a sphere dimer, from the electroquasistatic to the full-wave regimes.

In the quasistatic limit, we have established a $\parity\trev$-symmetric realization of an EPD, consisting of two spheres with complex-conjugate susceptibilities. Beyond this quasistatic limit, retardation breaks $\parity\trev$-symmetry; yet \textit{real-frequency} EPDs can still be attained by appropriate choices of the dispersion of the materials of the two spheres and geometry. Casting the problem as an equivalent two-port circuit yields analytical criteria for EPDs; the resulting predictions, benchmarked against rigorous multiparticle Mie theory for Drude metals, quantitatively capture the evolution and coalescence of the dimer resonances.

We have characterized the scattering, extinction, and absorption efficiencies across the EPD transition.   Close to an EPD, a single-parameter perturbation splits a resonance in the power spectrum into two efficiency peaks whose separation follows the characteristic square-root law seen for the eigenfrequencies.
This square-root splitting is a hallmark of EPDs and points to enhanced-sensitivity nanophotonic sensors based on minimal building blocks, with tunability via gain/loss control, plasma-frequency engineering, and interparticle spacing. Achieving the EPD condition exactly remains challenging, particularly at optical frequencies. By contrast, its implementation may be more practical in the microwave regime, where the two Drude permittivities can be implemented by metamaterials, with the active element realized through a tunable negative-resistance load (see, e.g., \cite{Ye_microwave_2014}). The theory can also be applied to anisotropic polarizable particles at microwave frequencies, like dipole antennas connected to an active negative-resistance component.

The framework is readily extensible: higher multipolar orders occurring at larger size parameters enter through additional mode impedances without altering the synthesis workflow; more complex dispersion can be incorporated via refined material impedances. We anticipate that the same methodology will generalize to complex arrays of scatterers, providing a systematic route to engineer, analyze, and exploit EPDs in realistic photonic platforms.

\begin{acknowledgements}
This work was supported by the Italian Ministry of University and Research under the PRIN-2022, Grant Number 2022Y53F3X ``Inverse Design of High-Performance Large-Scale Metalenses".
\end{acknowledgements}
%%%%%%%%%%%
\appendix
\section{NOTATIONS}
\label{sec:notations}
Let \(\mathbf{F} = \mathbf{F}(\mathbf{r})\) be a vector field defined on $\mathcal{B} = \mathcal{B}_1 \cup \mathcal{B}_2$.  We define the restriction of \(\mathbf{F}\) to the domain \(\mathcal{B}_\partm\) as 
\begin{equation}   
\mathbf{F}_{\partm}(\mathbf{r}) := 
\begin{cases}
    \mathbf{F}(\mathbf{r}) & \mathbf{r} \in \mathcal{B}_\partm, \\
    \mathbf{0} & \mathbf{r} \notin \mathcal{B}_\partm.
\end{cases}
\end{equation}
Throughout the paper, we use the standard inner product on \(\mathcal{B}\), defined by
\begin{equation}
\langle \mathbf{u}| \mathbf{v} \rangle := \frac{1}{\volume} \int_{\mathcal{B}} \mathbf{u}^*\cdot \mathbf{v} \dV,
\label{eq:InnerDef}
\end{equation}
where \(^*\) denotes complex conjugation and $\volume = 4\pi a^3/3$ is the volume of the particle.

The mean of the vector field \(\mathbf{F}\) over \(\mathcal{B}\) is defined as
\begin{equation}
\mean{\mathbf{F}} := \frac{1}{\volume} \int_{\mathcal{B}} \mathbf{F}(\mathbf{r}) \dV.
\label{eq:MeanDef}
\end{equation}

\section{EIGENVALUE PROBLEM}
The operator \(\mathscr{L}\) in Eq. \eqref{eq:EigProb} is defined as
\begin{multline}
    \mathscr{L}\{ \mathbf{J} \}(\mathbf{r}) := 
    \frac{\zeta_0}{j k_0 a} \nabla \oint_{\partial \mathcal{B}} g(\mathbf{r} - \mathbf{r}') \, \mathbf{J}(\mathbf{r}') \cdot \hat{\mathbf{n}}(\mathbf{r}') \, \mathrm{d}S' \\
    + j k_0 \zeta_0 \frac{1}{a} \int_{\mathcal{B}} g(\mathbf{r} - \mathbf{r}') \, \mathbf{J}(\mathbf{r}') \, \mathrm{d}\Omega',
    \label{eq:opLdef}
\end{multline}
where $\hat{\mathbf{n}}$ is the outward-pointing unit normal vector on $\partial \mathcal{B}$, and
\begin{equation}
    g(\mathbf{r} - \mathbf{r}') = \frac{e^{-j k_0 |\mathbf{r} - \mathbf{r}'|}}{4\pi |\mathbf{r} - \mathbf{r}'|}
\end{equation}
is the free-space Green’s function. It is rewritten as
\begin{equation}
\mathscr{L}\{ \vec{J} \} ({\vec{r}}) = \frac{\zeta_0}{a}jk_0 \int_{\mathcal{B}} \dyad{G}(\vec{r},\vec{r}') \cdot \vec{J}(\vec{r}')  \dV',
\label{eq:opLdef2}
\end{equation}
where 
%$\hat{\mathbf{n}}_\partm$ is the outward-pointing unit normal vector on $\partial \mathcal{B}_\partm$, 
\begin{equation}
\dyad{G}(\vec{r}, \vec{r}') := \left( \dyad{I} - k_0^{-2} \nabla \nabla' \right) g(\vec{r} - \vec{r}'),
\label{eq:dyadG}
\end{equation}
is the  free-space dyadic Green’s function, and 
$\dyad{I}$ the identity dyad. 

\section{COUPLED-DIPOLE APPROXIMATION}
\label{sec:CDA}
 In this section, we demonstrate the equivalence of our circuit model to the coupled-dipole approximation model \cite{yang_discrete_1995,park_surface-plasmon_2004, kelly_optical_2003, campione_complex_2011, steshenko_single_2009}.  

The full-wave integral equation \eqref{eq:constrel} can be significantly simplified by assuming that the electric field $\vec{E}_\partm$ is approximately constant within the domain $\mathcal{B}_\partm$ of the $\partm$-th particle, namely $\vec{E}_\partm(\vec{r})\approx \vec{E}_\partm(\vec{r}_\partm)$. Applying the mean operator \eqref{eq:MeanDef} to Eq. \eqref{eq:constrel} yields
\begin{equation}
a  Z^{\mat}_{\partm} \mean{\vec{J}_{\partm}} = \mean{\vec{E}_{\partm}} \approx \vec{E}_\partm(\vec{r}_\partm).
\label{eq:constrel_mean1}
\end{equation}
From Eq. \eqref{eq:Esca_s}, the total electric field $\vec{E}_\partm$ at the center of particle $\partm$ can be written as
\begin{equation}
\vec{E}_\partm(\vec{r}_\partm)= \vec{E}_\partm^\inc(\vec{r}_\partm) + \sum_{\nu=1}^2 \vec{E}^{\sca}_{\partm,\partn}(\vec{r}_\partm).
\end{equation}
The value of the electric field $\vec{E}^{\sca}_{\partm ,\partm}$ at the center of the $\partm$-th particle generated by the $\partm$-th particle itself is given by
\begin{equation}
\vec{E}^{\sca}_{\partm, \partm}(\vec{r}_\partm) \approx \mean{\vec{E}^{\sca}_{\partm,\partm}} = - \frac{1}{a} \sum_{h=1}^{\infty} \mathcal{Z}^h I^h_\partm \mean{\vec{j}^h_\partm}.
\end{equation}
In the limit of small and well-separated particles, and assuming that the real part of their permittivity is either negative or, if positive, not excessively large, the dominant contribution arises from the electric-dipole modes, each with impedance $\mathcal{Z}^h = \mathcal{Z}^{\mathrm{ed}}$ ($h=1,2,3$). This yields the approximation:
\begin{equation}
\mean{\vec{E}^{\sca}_{\partm,\partm}} \approx  -a \mathcal{Z}^{\mathrm{ed}} \mean{\vec{J}_\partm}.
\label{eq:Es_uu_2}
\end{equation}
According to Eq.~\eqref{eq:Esca_opL}, the contribution to the scattered field $\vec{E}^{\sca}_{\partm,\partn}(\vec{r}_\partm)$ due to the $\partn$-th particle alone and evaluated at the center of particle $\partm$ is 
$
\vec{E}^{\sca}_{\partm,\partn}(\vec{r}_\partm) = -a  \mathscr{L}_\partm \{\vec{J}_\partn\}(\vec{r}_\partm).
%\label{eq:Esca_opL2}
$
Since $|\vec{r}' - \vec{r}_\partn| \leq a$, if $a/d \ll 1$, we obtain
$
\dyad{G}(\vec{r}_\partm, \vec{r}') \approx \dyad{G}(\vec{r}_\partm, \vec{r}_\partn) = \dyad{G}_{\partm\partn}
%\label{eq:dyadG_center_center}
$, where
\begin{subequations}
  \begin{align}
   &\dyad{G}_{\partm \partn} = k_0 \left[ (1+jk_0d)\dyad{B}_{\partm\partn} - (k_0d)^2 \dyad{C}_{\partm\partn}\right]\frac{e^{-jk_0d}}{4\pi (k_0d)^3} \\
    & \dyad{B}_{\partm \partn} = 3 \vers{r}_\partm \vers{r}_\partn - \dyad{I} = 2 \vers{x}\vers{x} - \vers{y}\vers{y} - \vers{z}\vers{z}, \\
    & \dyad{C}_{\partm \partn} =  \vers{r}_\partm \vers{r}_\partn - \dyad{I} = - \vers{y}\vers{y} - \vers{z}\vers{z}.
\end{align} 
    \label{eq:Guv}
\end{subequations}
Using this approximation in Eq. \eqref{eq:opLdef2}, we find
\begin{equation}
\vec{E}_{\partm,\partn}^\sca(\vec{r}_\partm) \approx -\volume \zeta_0 jk_0 \dyad{G}_{\partm\partn} \cdot\mean{\vec{J}_\partn}.
\label{eq:Es_uv_2}
\end{equation}
Substituting Eqs. \eqref{eq:Es_uu_2} and \eqref{eq:Es_uv_2} into Eq. \eqref{eq:constrel_mean1}, we obtain
\begin{equation}
a(Z^{\mat}_{\partm} + \mathcal{Z}_{\mathrm{ed}})\mean{\vec{J}_\partm} + \volume \zeta_0 jk_0 \dyad{G}_{\partm\partn} \cdot \mean{\vec{J}_\partn} = \vec{E}_{\partm}^{\inc}(\vec{r}_{\partm}).
\label{eq:CDA_J}
\end{equation}
The average current $\mean{\vec{J}_\partm}$ is related to the electric-dipole moment $\vec{p}_\partm$ of particle $\partm$ as
\begin{equation}
\vec{p}_\partm := \volume\varepsilon_0 \zeta_0 \frac{\mean{\vec{J}_\partm}}{jk_0}.
\label{eq:Pdef}
\end{equation}
Using this relation, we arrive at the \textit{coupled-dipole equation}:
\begin{equation}
\alpha_\partm^{-1} \vec{p}_\partm - k_0^2 \sum_{\partn \neq \partm} \dyad{G}_{\partm\partn} \cdot\vec{p}_\partn = \varepsilon_0 \vec{E}_{\partm}^\inc(\vec{r}_\partm),
\label{eq:CDA}
\end{equation}
where $\alpha_\partm =\alpha_\partm(\omega)$ is the electric polarizability of the $\partm$-th particle, defined as
\begin{equation}
\alpha_\partm := \frac{\zeta_0 \volume}{jk_0a(Z^{\mat}_{\partm} + \mathcal{Z}^{\mathrm{ed}})} = \volume \left( \frac{1}{\chi_\partm} + jk_0a \frac{\mathcal{Z}^{\mathrm{ed}}}{\zeta_0} \right)^{-1} .
\label{eq:poldef}
\end{equation}
Notably, using the RC approximation of $\mathcal{Z}^\ed$ (see Eq. \eqref{eq:Zed_approx},  with $\mathcal{X}^{\ed}$ truncated at the order $\omega^{-1}$) in Eq. \eqref{eq:poldef} returns the Clausius-Mossotti polarizability with the radiative correction \cite{yang_discrete_1995,kelly_optical_2003, steshenko_single_2009}
\begin{equation}
    \alpha_\partm \approx 4\pi a^3 \left[\frac{\chi_\partm +3}{\chi_{\partm}} + j \frac{2}{3} (k_0a)^3\right]^{-1}.
    \label{eq:PolarizabilityMLWA}
\end{equation}
Accordingly, the proposed circuit model is equivalent to the standard CDA model, but with the polarizability in \eqref{eq:poldef} replacing the Clausius--Mossotti or long-wavelength-approximation polarizability typically adopted in CDA \cite{yang_discrete_1995,kelly_optical_2003,steshenko_single_2009}. Nevertheless, Eq. \eqref{eq:poldef} also reveals a key distinction between the two formulations. In the standard CDA, material properties, scattering properties, geometry, and frequency dependence are all incorporated into a single effective polarizability. In contrast, the eigenvalue formulation in Eq. \eqref{eq:EigProb} isolates the material contribution from the geometric and radiative terms, which substantially simplifies the derivation of the EPD conditions and adds physical insight.

\section{ASYMPTOTIC EXPANSIONS}
\label{sec:Asymptotic}
The impedance of the electric-dipole mode admits a Laurent series expansion around $k_0a = 0$. Accordingly, the radiation resistance $\mathcal{R}^{\text{ed}}$ and the modal reactance $\mathcal{X}^{\text{ed}}$ of the electric-dipole mode of a sphere can be written as \cite{forestiere_first-principles_2024}
\begin{subequations}
\begin{align}
    \label{eq:Red_approx}
    \mathcal{R}^{\ed} &= \frac{2}{9}\zeta_0 (k_0a)^2 + o\left((k_0a)^3\right), \\
    \label{eq:Xed_approx}
    \mathcal{X}^{\ed} &= -\frac{1}{\omega C^{\text{ed}}} + \omega L^{\mathrm{ed}} + o\left((k_0a)^2\right),
\end{align}
\label{eq:Zed_approx}
\end{subequations}
where $o(\cdot)$ denotes the Bachmann–Landau little-o, and
\begin{equation}
C^{\ed} = 3\varepsilon_0 a,  \hspace{3ex} L^{\ed} = \frac{4}{15} \mu_0 a.
    \label{eq:ED_lowfreq_param}
\end{equation}
Eqs.~\eqref{eq:Zed_approx} show that, in the small-particle limit $k_0a\ll1$, the impedance $\mathcal{Z}^{\mathrm{ed}}$ can be approximated by that of a series RLC circuit.

{

The mutual resistance $R_{\mutua}^h = \Re\{Z_{\mutua}^h\}$ and reactance $X_{\mutua}^h = \Im\{Z_{\mutua}^h\}$ are

\begin{subequations}
    \begin{align}
    R_{\mutua}^x &= \frac{2}{3}\zeta_0 \rho^2\,\left(\frac{\sin(k_0d)}{k_0d} - \cos (k_0d)\right); \label{eq:Rx_12}\\
 X_{\mutua}^x &= \frac{2}{3}\zeta_0 \rho^2\,\left(\frac{\cos(k_0d)}{k_0d} + \sin (k_0d)\right); \label{eq:Xx_12} \\
 R_{\mutua}^{\overset{y}{z}} &= \frac{1}{3}\zeta_0 \rho^2\,\left([(k_0d)^2-1]\frac{\sin(k_0d)}{k_0d} + \cos (k_0d)\right); \label{eq:Ry_12}\\
 X_{\mutua}^{\overset{y}{z}} &= \frac{1}{3}\zeta_0 \rho^2\,\left([(k_0d)^2-1]\frac{\cos(k_0d)}{k_0d} - \sin (k_0d)\right). \label{eq:Xy_12}
 % \label{eq:Lt}
\end{align}
\label{eq:RXmutua}
\end{subequations}

By expanding $Z_{\mutua}^h$ in Laurent series around $k_0d = 0$ we obtain
\begin{subequations}
\begin{align}
\label{eq:Rmu_approx}
&R_{\mutua}^h =   \frac{2}{9}\zeta_0 (k_0a)^2 + o\left((k_0d)^3\right),\\
\label{eq:Xmu_approx}
& X_{\mutua}^h = -\frac{1}{\omega C_{\mutua}^h} + \omega L_{\mutua}^h + o\left((k_0d)^2\right)
\end{align}
\label{eq:Zmu_approx}
\end{subequations}
where 

\begin{equation}
\frac{C_{\mutua}^h}{C^\ed} = \begin{cases}
    -\frac{1}{2}\rho^{-3} & \text{if } h=x,\\
    \rho^{-3} &\text{if }h=y,z
\end{cases}  \,  , \frac{L_{\mutua}^h}{L^\ed} = \begin{cases}
    \frac{5}{2}\rho &\text{if } h=x,\\
    \frac{15}{12} \rho  &\text{if } h=y,z,\\
\end{cases}
\label{eq:Zmu_low_freq_param}
\end{equation}
and $\rho = a/d$, is the geometric ratio. Eqs. \eqref{eq:Zmu_approx} show that, in the small-separation regime ($k_0d\ll1$), the mutual impedance is ohmic-capacitive for both longitudinal and transverse scenarios, as illustrated in Fig. \ref{fig:Z_mutua}(c)-(f).  
}

{
\section{STATE-SPACE EQUATIONS AND EPDs}
\label{sec:Jordan}
The dynamical behavior of the two-port circuit shown in Fig.~\ref{fig:2port}(a) can be described in the time domain using the standard state--space formalism
\begin{equation}
   \dot{\vec{x}} = \boldsymbol{A}\vec{x} + \boldsymbol{B}\vec{e},\qquad
   \vec{i} = \boldsymbol{C}\vec{x} + \boldsymbol{D}\vec{e},
\label{eq:SS2_time}
\end{equation}
where $\vec{x} = \vec{x}(t) \in \mathbb{R}^N$ is the state vector, $\vec{e}(t) = [e_1,e_2]^\intercal$ is the input vector, and $\vec{i}(t) = [i_1,i_2]^\intercal$ is the output vector. The matrix $\boldsymbol{A} \in \mathbb{R}^{N \times N}$ is the system (state) matrix, while
$\boldsymbol{B}$, $\boldsymbol{C}$, and $\boldsymbol{D}$ are the input, output, and feedthrough matrices. Since the circuit exhibits no instantaneous dependence of the output on the input, one has $\boldsymbol{D}=\boldsymbol{0}$.

Without loss of generality, we set  $e_2 =0$, as a consequence the circuit in Fig. \ref{fig:2port}(a) becomes the one shown in Fig.~\ref{fig:2port}(b), which is governed by the same state matrix $\boldsymbol{A}$.   The corresponding state--space equations read
\begin{equation}
   \dot{\vec{x}} = \boldsymbol{A}\vec{x} + \vec{b}\,e_1, \qquad
   i_1(t) = \vec{c}^{\intercal} \vec{x}.
\label{eq:SS_time}
\end{equation}
where $\vec{b}$ is the column vector corresponding to the first column of $\boldsymbol{B}$, and $\vec{c}$ is the column vector whose transpose $\vec{c}^{\intercal}$ corresponds to the first row of $\boldsymbol{C}$.
\begin{figure}[t]
    \centering
    \includegraphics[width=0.65\linewidth]{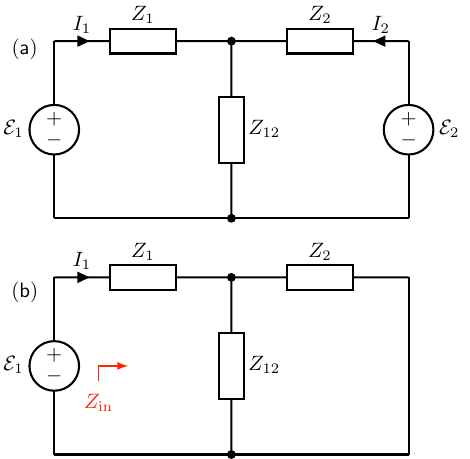}
    \caption{(a) Two-port circuit in the Laplace domain. Here, $\mathcal{E}_{1,2} = \mathcal{L}\{e_{1,2}\}$ and $I_{1,2} = \mathcal{L}\{i_{1,2}\}$, where $\mathcal{L}$ denotes the Laplace Transform. (b) Definition of the input impedance $Z_{\mathrm{in}}$}
    \label{fig:2port}
\end{figure}

Without loss of generality, we assume zero initial conditions, i.e., $\mathbf{x}(0)=0$. Applying the Laplace transform to~\eqref{eq:SS_time},
one obtains the input--output relation
\begin{equation}
I_1(s) = Y(s)\,\mathcal{E}_1(s),
\qquad
Y(s) = \mathbf{c}^\intercal (s\mathbb{I}-\boldsymbol{A})^{-1} \mathbf{b}.
\label{eq:TF}
\end{equation}
Here,  $I_1$ and $\mathcal{E}_1$ denote the Laplace transforms of $i_1$ and $e_1$, respectively; $s = \sigma + j \omega$ denotes the Laplace variable, and $\mathbb{I}$ is the identity matrix. The function $Y(s)$ is the \textit{transfer function} of the system.

For a linear two-port network, the input impedance is defined as \cite{chua_linear_1987} [see Fig. \ref{fig:2port}(b)]
\begin{equation}
    Z_{\mathrm{in}}(s)
    = \left. \frac{\mathcal{E}_1(s)}{I_1(s)} \right|_{\mathcal{E}_2 = 0},
\end{equation}
so that $Y(s) = Z_{\mathrm{in}}^{-1}(s)$. According to Eq. ~\eqref{eq:TF}, if $\lambda_i$ is a pole of \( Y(s) \) (equivalently, a zero of \( Z_{\mathrm{in}}(s) \)), then $\lambda_i$ is an eigenvalue of \( \boldsymbol{A} \).

Moreover, if \( Y(s) \) has a pole of order \( n_i \) at \( s = \lambda_i \), then the Jordan canonical form of $ \boldsymbol{A}$ contains a Jordan block associated with $\lambda_i$ of dimension at least $n_i$. For a \textit{minimal realization}, i.e., in the absence of pole-zero cancellations, the dimension of the largest Jordan block associated with $\lambda_i$ is exactly $n_i$ \cite{kailath_linear_1980}.

As a direct consequence, the presence of higher-order poles $(n_i \ge 2)$ in the transfer function $Y(s)=1/Z_{\mathrm{in}}(s)$ implies that the state matrix $ \boldsymbol{A}$ is non-diagonalizable. Such spectral degeneracies correspond to exceptional point degeneracies in the dynamical description of the circuit.
}
   
\section{PARITY-TIME TRANSFORMATION}
\label{sec:App_PT_symmetry}

The \textit{parity} (\(\parity\)) and \textit{time-reversal} (\(\trev\)) operators acting on \(\boldsymbol{\mathcal{Z}}\) are defined as
\begin{subequations}
\begin{align}
    \parity: \quad &\boldsymbol{\mathcal{Z}}(\omega,\param) \mapsto 
    \parity\{\boldsymbol{\mathcal{Z}}\} := 
    \mathbb{P}\,\boldsymbol{\mathcal{Z}}(\omega,\param)\,\mathbb{P}^\intercal, \\
    \trev: \quad &\boldsymbol{\mathcal{Z}}(\omega,\param) \mapsto 
    \trev\{\boldsymbol{\mathcal{Z}}\} := 
    -\boldsymbol{\mathcal{Z}}^*(\omega^*, \param),
\end{align}
\label{eq:PTop_def}
\end{subequations}
where \(\mathbb{P}\) denotes the $2 \times 2$ permutation matrix with entries 
\([\mathbb{P}]_{\partm \partn} = 1 - \delta_{\partm \partn}\).

A system is said to be \(\PT\)-\textit{symmetric} if it is invariant under the combined action of these two operators \cite{bender_real_1998}, i.e. $
    \PT\{\boldsymbol{\mathcal{Z}}\} = \boldsymbol{\mathcal{Z}}.
$

We now show that the \(\PT\)-symmetry implies condition~\eqref{eq:FconjSimm}. Let $F(\omega) := \det\!\bigl(\boldsymbol{\mathcal{Z}}(\omega)\bigr)$. Under the action of the \(\PT\)-operator, the impedance matrix transforms as: $ \PT\{\boldsymbol{\mathcal{Z}}(\omega)\}
    = -\,\mathbb{P}\,\boldsymbol{\mathcal{Z}}^{*}(\omega^*)\,\mathbb{P}^{\intercal}$. Taking determinants and using the \(\PT\)-invariance of the system yields $F(\omega;\param)
    = \det\!\left(-\mathbb{P}\,\boldsymbol{\mathcal{Z}}^{*}(\omega^*)\,\mathbb{P}^{\intercal}\right)$. Since \(\det(\mathbb{P}) = -1\), we obtain $F(\omega;\param)
    = \det\!\bigl(\boldsymbol{\mathcal{Z}}(\omega^*)\bigr)^{*}
    = \bigl[F(\omega^*;\param)\bigr]^{*}$. Thus, $F(\omega^*;\param) = \bigl[F(\omega;\param)\bigr]^{*}$, which proves the claim.

\section{OPTICAL CROSS-SECTIONS}
\label{sec:ScatteringAppendix}
The time-averaged powers scattered ($P_\sca$), absorbed ($P_\abs$) and extinguished ($P_\ext$) by the dimer embedded in vacuum are \cite{bohren_absorption_1998}
\begin{subequations}
    \begin{align}
    &P_{\sca} = - \frac{V_a}{2}  \Re  \sum_{\partm = 1}^2\langle \vec{J}_\partm| \vec{E}^{\sca}_\partm \rangle,
    \label{eq:Psca2} \\
    & P_{\abs} =  \frac{V_a}{2}  \Re \sum_{\partm = 1}^2\langle \vec{J}_\partm| \vec{E}_\partm \rangle,
    \label{eq:Pabs1} \\
    &  P_{\ext} = \frac{V_a}{2}  \Re \sum_{\mu = 1}^2\langle \vec{J}_\partm| \vec{E}^{\inc}_\partm \rangle.
\end{align}
\end{subequations}
The normalization of these powers by $P_\inc = \pi a^2 |\vec{E}^{\inc}|^2/\zeta_0 $, defines the extinction ($Q_{\ext}$), scattering ($Q_{\sca}$), and absorption ($ Q_{\abs}$) efficiencies \cite{bohren_absorption_1998}.
In the CDA, these three powers can be rewritten as
\begin{subequations}
\begin{align}
    &     P_{\sca} \approx \frac{1}{2} \Re \sum_{h = 1}^3 \vec{I}^{h\dagger} \begin{bmatrix}
        \mathcal Z^{\ed} &Z_M^h \\
        Z_M^h & \mathcal Z^{\ed}
    \end{bmatrix} \vec{I}^h, \\
    &     P_{\abs}  \approx \frac{1}{2} \Re \sum_{h = 1}^3 \vec{I}^{h\dagger} \begin{bmatrix}
        Z_1^\mat &0 \\
        0 & Z_2^\mat
    \end{bmatrix} \vec{I}^h, \\
        & P_{\ext} \approx \frac{1}{2}\Re \sum_{h=1}^3 \vec{I}^{h\dagger} \boldsymbol{\mathcal{E}}^h,
\end{align}
\label{eq:CrossSec}
\end{subequations}
where $\dagger$ denotes the conjugate transpose.

{\section{PERTURBATION ANALYSIS}
\label{sec:Sensitivity}

Consider a dimer operating at an EPD, characterized by the pair \( (\omega_\ep, \param_\ep) \). We introduce a perturbation of the form \( \delta \param_i  =  \param - \param_\ep \), which affects only the \( i \)-th component of the parameter vector $\param_\ep$. Our objective is to analyze the sensitivity of the system response, in particular the scattered power, with respect to variations in the parameter \( \vartheta_i \).

Near an EPD, the eigenfrequencies admit a Puiseux series expansion \cite{welters_explicit_2011} of the form
\begin{equation}
    \omega_{\pm}(\vartheta_i) = \omega_{\ep} \pm j b \sqrt{\delta \vartheta_i} + o(\sqrt{\delta \vartheta_i}),
    \label{eq:Puiseux}
\end{equation}
where \( \omega_{\ep} := \omega(\vartheta_{\ep}) \), and $b$ is a nonzero constant.

An immediate implication of Eq.~\eqref{eq:Puiseux} for the power spectrum is that the resonance frequency, 
\(\omega_{\mathrm{r}\pm} = \Re \{\omega_{\pm}\}\), is highly sensitive to perturbations. In particular, the resonance frequency splitting, $\Delta \omega_\mathrm{r} = \omega_{\mathrm{r+}} - \omega_{\mathrm{r-}}$, is given by
\begin{equation}
\Delta \omega_\mathrm{r} = 
\begin{cases}
    -2 \sqrt{\delta \vartheta_i} \, \Im \{b\}, & \delta \vartheta_i \geq 0, \\[6pt]
    -2 \sqrt{|\delta \vartheta_i|} \, \Re \{b\}, & \delta \vartheta_i < 0.
\end{cases}   
\end{equation}

Finally, we define the sensitivity function \( S_{\vartheta_i} \) as
\begin{equation}
    S_{\vartheta_i}\{\Delta \omega_\mathrm{r}\} := 
    \left| \frac{\partial \Delta \omega_\mathrm{r}}{\partial \vartheta_i} \right|.
\label{eq:sensitivity_def}
\end{equation}
Due to the square-root dependence, the sensitivity diverges asymptotically as $
S_{\vartheta_i} \sim |\delta \vartheta_i|^{-1/2}$, for $\delta \vartheta_i \to 0$.

\section{TRANSVERSE POLARIZATION}
\label{sec:tranv}

In Sec. \ref{sec:results}, results are presented for the longitudinal polarization of the dipolar mode ($h=x$). In this appendix, we discuss the transverse polarization case ($h=y,z$).

The primary difference arises from the coupling mechanism between the dipoles. Within the circuit-based formalism, this modifies only the mutual impedance $Z_M$. This difference becomes particularly transparent in the quasistatic limit. In this regime, Eq. \eqref{eq:wstatic_circuit} becomes
\begin{equation}
\left(\frac{\omega}{\omega_0}\right)^4 
+ \left( \hat{\gamma}^2 - 2 \right)\left(\frac{\omega}{\omega_0}\right)^2 
+ 1 - \rho^6 = 0,
\label{eq:wstatic_transverse}
\end{equation}
The corresponding eigenfrequencies are
\begin{equation}
\omega_{\pm}(\hat \gamma,\rho) = \omega_0 
\sqrt{1 - \frac{\hat{\gamma}^2}{2} 
\pm \frac{1}{2} 
\sqrt{\hat{\gamma}^4 - 4\hat{\gamma}^2 + 4\rho^6}}.
\label{eq:w_12_transverse}
\end{equation}
An EPD occurs when the inner discriminant vanishes:
\begin{equation}
\hat{\gamma}^4 - 4\hat{\gamma}^2 + 4 \rho^6 = 0,
\label{eq:gamma_epd_eq_tran}
\end{equation}
which is quadratic in $\hat{\gamma}^2$. The corresponding discriminant is 
$\Delta = 16(1 - \rho^6)$, which remains strictly positive under the 
non-overlapping condition $d > 2a$ (i.e., $\rho < 1/2$). 
Therefore, as in the longitudinal case, two real solutions for 
$\hat{\gamma}^2$ exist throughout the 
physically admissible parameter range.

Compared to \eqref{eq:gamma_epd_eq}, Eq. \eqref{eq:gamma_epd_eq_tran} differs only in the coefficient of the $\rho^6$ term.

%\bibliographystyle{ieeetr}
%\bibliography{references,references_emanuele_zotero,references_2}

\end{document}